\documentclass[twocolumn,showpacs,amsmath,floatfix,prl,aps]{revtex4}
\usepackage{graphicx}

\begin{document}
\title{First-principles approach to Non-Collinear Magnetism : towards Spin Dynamics}
\author{S. Sharma$^{1,2,5}$}
\email{sangeeta.sharma@uni-graz.at}
\author{J. K. Dewhurst$^{2,3}$}
\author{C. Ambrosch-Draxl$^{2,4}$}
\author{S. Kurth$^5$}
\author{N. Helbig$^{1,5}$}
\author{S. Pittalis$^5$}
\author{S. Shallcross$^6$}
\author{L. Nordstr\"om$^7$}
\author{E. K. U. Gross$^5$}
\affiliation{1 Fritz Haber Institute of the Max Planck Society, Faradayweg 4-6, D-14195 Berlin, Germany.}
\affiliation{2 Institut f\"{u}r Physik, Karl--Franzens--Universit\"at Graz,
Universit\"atsplatz 5, A--8010 Graz, Austria.}
\affiliation{3 School of Chemistry, The University of Edinburgh, Edinburgh
EH9 3JJ}
\affiliation{4 Department Materials Physics, University of Leoben, 
Erzherzog Johann - Strasse 3, A-8700 Leoben},
\affiliation{5 Institut f\"{u}r Theoretische Physik, Freie Universit\"at Berlin,
 Arnimallee 14, D-14195 Berlin, Germany.}
\affiliation{6 Department of Physics, Technical University of Denmark
DK-2800 Lyngby, Denmark.}
\affiliation{7 Department of Physics, Uppsala University, Box 530, 751 21 Uppsala, 
Sweden.}

\date{\today}

\begin{abstract}
A description of non-collinear magnetism in the framework of spin-density 
functional theory is presented for the exact exchange energy functional which 
depends explicitly on two-component spinor orbitals. The equations for the effective 
Kohn-Sham scalar potential and magnetic field are derived within the optimized 
effective potential (OEP) framework. With the example of a magnetically frustrated Cr
monolayer it is shown that the resulting magnetization density exhibits much more 
non-collinear structure than standard calculations. Furthermore, a 
time-dependent  generalization of the non-collinear OEP method is well suited 
for an {\it ab-initio} description of spin dynamics. We also show that the magnetic 
moments of solids Fe, Co and Ni are well reproduced. 
\end{abstract}


\maketitle

The extension of the 
original density functional theory (DFT) Hohenberg-Kohn-Sham approach to the case 
of spin polarized systems was given, under the name spin DFT (SDFT) more than 
three decades ago \cite{bath72}. While this formulation was for arbitrary directions 
of the magnetization vector field, even today most applications are based on a 
restricted collinear version. This has the advantage of computational simplicity: 
one then works with two separate KS equations, one yielding the
spin-up orbitals the other the spin-down orbitals, whereas the
general formulation involves Pauli spinors.
Nevertheless, there exists a wealth of non-collinearity in nature. To give
only a few examples it is widely seen in molecular magnets, exchange frustrated
solids ($\gamma$-Fe, spin glasses), and all magnets at finite temperatures.

Crucial for practical calculations using SDFT is the approximation made for the 
exchange-correlation (xc) energy functional. The Local Spin Density 
Approximation (LSDA) and the Generalized Gradient Approximations (GGAs) are 
currently the most popular ones. These have been 
developed for collinear magnetism, and their use in non-collinear situations
relies on the magnetization, ${\bf m}({\bf r})$, and exchange correlation 
magnetic field, ${\bf B}_{\rm xc}({\bf r})$, being made collinear in a local 
reference frame at each point in space \cite{kub88}. This is only possible with 
purely local functionals like LSDA \cite{lars96,oda98}, though 
it has been used under additional approximations for gradient functionals 
as well \cite{lars02}. 
Such approximations (that lead to locally collinear magnetization and xc 
magnetic field) cause ${\bf m}({\bf r}) \times {\bf B}_{\rm xc}({\bf r})$ to 
vanish everywhere in space.
As noted recently, this fact renders the adiabatic time dependent extension of
these functionals improper \cite{capelle01} for the study of spin dynamics, 
because in the absence of external magnetic fields and within adiabatic 
approximation, the local torque on the spins 
(${\bf m}({\bf r,t}) \times {\bf B}_{\rm xc}({\bf r,t})$) vanishes \cite{capelle03}.  
This is a serious limitation since the dynamics
of the spin degree of freedom is responsible for a number of important phenomena 
such as spin injection, the dynamics of Bloch walls, spin wave excitations
\cite{gebauer00}, and spin filtering, mechanisms crucial for recent developments
in spintronics \cite{wolf}.
The search for approximate xc functionals which depend on all three components
of the spin magnetization {\bf m} beyond the form of the locally collinear LSDA
has remained a major challenge in the description of non-collinear magnetism.

In recent years, an alternative route to the construction of approximate xc
functionals has enjoyed increasing interest. These involve functionals depending
explicitly on the single-particle KS orbitals which, through the KS single-particle 
equation, are \emph{implicit} functionals of the density \cite{angel06}. Technically, one needs 
to employ the Optimized Effective Potential (OEP) \cite{tal76} method to compute 
the local xc potential. The simplest orbital-dependent approximation to the xc energy 
is the EXact eXchange (EXX) functional which is the Fock exchange energy but evaluated 
with KS orbitals (i.e. orbitals coming from a local potential). A number of 
successful EXX calculations have been reported for 
semiconductors \cite{stadele99,magyar04,sharma05} and magnetic metals \cite{kot98}. 
However, for magnetic systems again the collinear formalism has been employed.

In this Letter we extend the OEP formalism for SDFT to non-collinear magnetic systems.
Most importantly, we do not rely on a condition of local collinearity and
treat the wavefunctions as Pauli spinors for high lying and Dirac spinors for 
deep lying (3 Ha below the Fermi level) electrons.
Using the EXX functional, we demonstrate with the 
example of an unsupported Cr(111) monolayer, that 
(i) the magnetization and ${\bf B}_{\rm xc}$ are generally not locally 
parallel in contrast to what has been assumed in all calculations to date and
(ii)  that the non-collinearity is much more pronounced than found with the
LSDA functional. Against popular belief \cite{kurz04}, we 
find that this non-collinearity 
is not restricted to just the interstitial region but spreads all the way to 
the atom center. With the examples of bulk Fe, Co and Ni we further show that our 
formalism can also be effectively used for collinear magnets.

To derive the OEP equations in the general non-collinear case, we start with the 
Kohn-Sham (KS) equation for two-component spinors $\Phi_i$, which has 
the form of a Pauli equation. For non-interacting electrons moving in an effective
scalar potential $v_{\rm s}$ and a magnetic vector field 
${\bf B}_{\rm s}$ it reads as (atomic units are used throughout)
\begin{align}\label{eq:KS}
\left(-\frac{1}{2}\nabla^2+v_{\rm s}({\bf r})
+\mu_B \boldsymbol\sigma\cdot{\bf B}_{\rm s}({\bf r})
\right)\Phi_i({\bf r})=\varepsilon_i\Phi_i({\bf r}).
\end{align}
This equation can be derived by minimizing the total
energy which, in SDFT, is given as a functional of the density 
$\rho({\bf r})=\sum_{i}^{\rm occ} \Phi_i^\dag({\bf r})\Phi_i({\bf r})$ and the
magnetization density
${\bf m}({\bf r})
=\mu_B \sum_{i}^{\rm occ} \Phi_i^\dag({\bf r})\boldsymbol\sigma\Phi_i({\bf r})$. 
For a given external scalar potential $v_{\rm ext}$ and magnetic field 
${\bf B}_{\rm ext}$ this total energy reads
\begin{widetext}
\begin{eqnarray}\label{eq:Etot}
E[\rho,{\bf m}] &=& T_s[\rho,{\bf m}]+\int \rho({\bf r}) v_{\rm ext}({\bf r})\,d{\bf r}
+\int {\bf m}({\bf r})\cdot{\bf B}_{\rm ext}({\bf r})\,d{\bf r}
+U[\rho]
+E_{\rm xc}[\rho,{\bf m}]\\
&=&\nonumber
\sum_{i}^{\rm occ} \varepsilon_i
-\int \rho({\bf r}) v_{\rm xc}({\bf r})\,d{\bf r}
-\int {\bf m}({\bf r})\cdot{\bf B}_{\rm xc}({\bf r})\,d{\bf r}
-U[\rho]+E_{\rm xc}[\rho,{\bf m}],
\end{eqnarray}
\end{widetext}
where $U[\rho]=1/2\int\!\!\int\rho({\bf r})\rho({\bf r}')/|{\bf r}-{\bf r}'|
\,d{\bf r}\,d{\bf r'}$ is the Hartree energy. The xc potential and xc magnetic 
field are given by
\begin{equation}\label{eq:potentials}
v_{\rm xc}({\bf r})=\frac{\delta E_{\rm xc}[\rho,{\bf m}]}{\delta\rho({\bf r})} 
\quad {\rm and} \quad
{\bf B}_{\rm xc}({\bf r})=\frac{\delta E_{\rm xc}[\rho,{\bf m}]}{\delta{\bf m}({\bf r})},
\end{equation}
respectively. The exact functional form of $E_{\rm xc}[\rho,{\bf m}]$ is unknown 
and has to be approximated in practice. 

Assuming that the densities $(\rho,{\bf m})$ are non-interacting 
$(v,{\bf B})$-representable
one may, equivalently, minimize the total-energy functional (\ref{eq:Etot}) over 
the effective scalar potential and magnetic field. Thus the conditions
\begin{align}\label{eq:var1}
\left.\frac{\delta E[\rho,{\bf m}]}
{\delta v_{\rm s}({\bf r})}\right|_{{\bf B}_{\rm s}}=0
\qquad{\rm and}\qquad
\left.\frac{\delta E[\rho,{\bf m}]}
{\delta {\bf B}_{\rm s}({\bf r})}\right|_{v_{\rm s}}=0
\end{align}
must be satisfied.

If the functional derivatives in Eq. \ref{eq:var1} are evaluated for an xc 
functional that depends explicitly on the KS spinors, one obtains the
natural extension of the OEP equations to non-collinear magnetism.
By the usage of spinor valued wavefunctions we can stay within a single global 
reference frame, in contrast to the case where functionals originally designed 
for collinear magnetism are used in a non-collinear context by introducing a local 
reference frame at each point in space. The most commonly used orbital functional 
is the EXX energy given by
\begin{align}\label{eq:EXX}
E^{\rm EXX}_{\rm x}[\{\Phi_i\}]\equiv -\frac{1}{2}\int\int\sum_{i,j}^{\rm occ}\frac{\Phi_i^\dag({\bf r})
\Phi_j({\bf r})\Phi_j^\dag({\bf r}')\Phi_i({\bf r}')}{|{\bf r}-{\bf r}'|}
\,d{\bf r}\,d{\bf r}'
\end{align}
where the label occ indicates that the summation runs only over occupied states.
In the following we restrict ourselves to an \emph{exchange-only} treatment although 
generalization to other orbital functionals is straightforward.

For the energy functional Eq.(\ref{eq:Etot}) using the EXX approximation to
$E_{\rm xc}$ one obtains the following coupled integral equations for the exchange
potential and magnetic field
\begin{align}\label{RV}
R_v({\bf r})&\equiv\left.\frac{\delta E[\rho,{\bf m}]}
{\delta v_{\rm s}({\bf r})}\right|_{{\bf B}_{\rm s}}\nonumber\\
&=\sum_{i}^{\rm occ}\sum_{j}^{\rm un}\left(\Lambda_{ij}\frac{\rho_{ij}({\bf r})}
{\varepsilon_i-\varepsilon_j}+{\rm c.c.}\right)=0
\end{align}
and
\begin{align}\label{RB}
{\bf R}_{\bf B}({\bf r})&\equiv\left.\frac{\delta E[\rho,{\bf m}]}
{\delta {\bf B}_{\rm s}({\bf r})}\right|_{v_{\rm s}}\nonumber\\
&=\sum_{i}^{\rm occ}
\sum_{j}^{\rm un}\left(\Lambda_{ij}\frac{{\bf m}_{ij}({\bf r})}
{\varepsilon_i-\varepsilon_j}+{\rm c.c.}\right)=0,
\end{align}
where
$\rho_{ij}({\bf r})=\Phi_i^\dag({\bf r})\Phi_j({\bf r})$,
${\bf m}_{ij}({\bf r})=\mu_B \Phi_i^\dag({\bf r})\boldsymbol\sigma\Phi_j({\bf r})$ 
and $j$ runs only over the unoccupied states. The matrix
$\Lambda$ is given by
\begin{align}
\Lambda_{ij}=\left(V_{ij}^{\rm NL}\right)^\dag
-\int\rho_{ij}^\dag({\bf r})v_{\rm x}({\bf r})\,d{\bf r}
-\int{\bf m}_{ij}^\dag({\bf r})\cdot{\bf B}_{\rm x}({\bf r})\,d{\bf r},
\end{align}
where
\begin{align}\label{eq:VNL}
V_{ij}^{\rm NL}=-\sum_{k}^{\rm occ}\int\int\frac{\Phi_i^\dag({\bf r})
\Phi_k({\bf r})\Phi_k^\dag({\bf r}')\Phi_j({\bf r}')}{|{\bf r}-{\bf r}'|}
\,d{\bf r}\,d{\bf r}',
\end{align}
are the non-local matrix elements of the Coulomb interaction between states
$i$ and $j$.

To ensure that our numerical analysis be as accurate as possible, we use the 
full-potential linearized augmented plane wave (FP-LAPW) method \cite{singh} 
implemented within the EXCITING code \cite{exciting}. 
Here the single electron potential is calculated exactly without any shape 
approximation and the space is divided into muffin-tin (MT) regions, where atomic
orbitals are used as a basis and interstitial region, where  plane waves are
used as a basis. The deep lying core states (3 Ha below the Fermi level) are 
treated as Dirac spinors and valence states as Pauli spinors. More importantly 
the magnetization density and xc magnetic field are both treated as unconstrained 
vector fields throughout space.
In our implementation of the OEP method the exchange fields are iteratively 
updated by subtracting the residue functions $R_v$ and ${\bf R}_{\bf B}$ from 
the exchange fields. In other words, if $i$ is the iteration number then
\begin{align}
v_{\rm x}^i({\bf r})=v_{\rm x}^{i-1}({\bf r})-\tau R_v^i({\bf r}),\nonumber\\
{\bf B}_{\rm x}^i({\bf r})={\bf B}_{\rm x}^{i-1}({\bf r})
-\tau {\bf R}_{\bf B}^i({\bf r})
\end{align}
is repeated until convergence is reached, with $R_v^i$ and ${\bf R}_{\bf B}^i$
calculated by inserting $v_{\rm x}^{i-1}$ and ${\bf B}_{\rm x}^{i-1}$ into
Eqs. (\ref{RV}) and (\ref{RB}). $\tau$ is the mixing chosen in such a manner 
as to achieve a speedy convergence.
In the collinear case this method is similar to the one previously suggested in 
the Ref. \onlinecite{kum03}.

\begin{figure}[ht]
\centerline{\includegraphics[width=\columnwidth,angle=0]{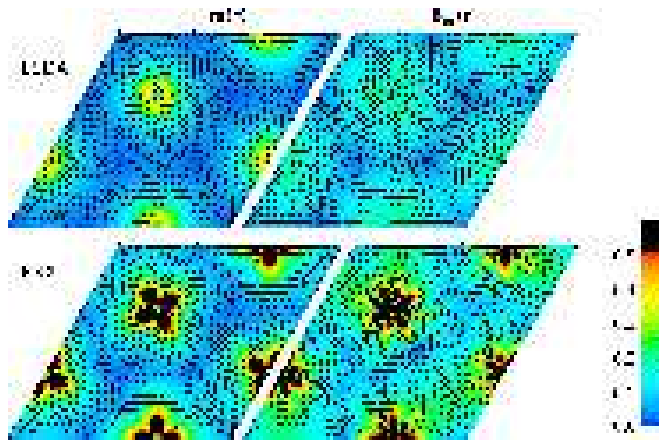}}
\caption{\label{fig:Cr1} Fully non-collinear magnetization density and {\bf B}
field obtained using the LSDA and exchange-only EXX functionals for an unsupported
Cr-monolayer in N\'eel state. Arrows indicate the direction and information about
the magnitude (in atomic units) is given in the colour bar.}
\end{figure}
In order to explore the impact of treating non-collinear magnetism in the
way outlined above we compare our approach with the standard LSDA functional using 
the example of an unsupported Cr (111) monolayer.
We set the lattice parameter of the Cr-monolayer to that of the Ag (111) surface. 
The result is a topologically frustrated anti-ferromagnet, known from LSDA calculations to 
exist as a non-collinear N\'eel state with the net magnetization direction of the three
non-equivalent atoms pointing at 120$^\circ$ to each other. In Fig. \ref{fig:Cr1} we
show the magnetization density and {\bf B} field for both the LSDA and EXX functionals. 
Both find, as they must, the non-collinear N\'eel state, and in fact the EXX and LSDA MT 
averaged moments are similar, being 2.60 $\mu_B$ and 2.0 $\mu_B$, respectively.
The details of the xc density and field however are very different with the EXX 
functional producing a lot more structure, in contrast to its fairly homogeneous 
LSDA counterpart. In the past, the LSDA results (of the kind 
shown in Fig. 1), which show almost no non-collinearity in the MT region, led to
the conclusion that it is sufficient to treat only the interstitial region as
non-collinear \cite{kurz04}. The present work shows that orbital functionals such 
as EXX are more sensitive to the atomic shell structure and this sensitivity 
also manifests itself in the magnetization density and exchange {\bf B} field.
This is clear from the flower petal like structure visible in the magnitude of 
EXX density and {\bf B} field.
The N\'eel walls are also much narrower in the EXX case. Adding LSDA correlations
to the EXX functional does not significantly change these results.
A striking feature of the EXX {\bf B} field  is that, unlike its LSDA
counterpart, it is not locally parallel to the magnetization density.


\begin{figure}[ht]
\centerline{\includegraphics[width=\columnwidth,angle=0]{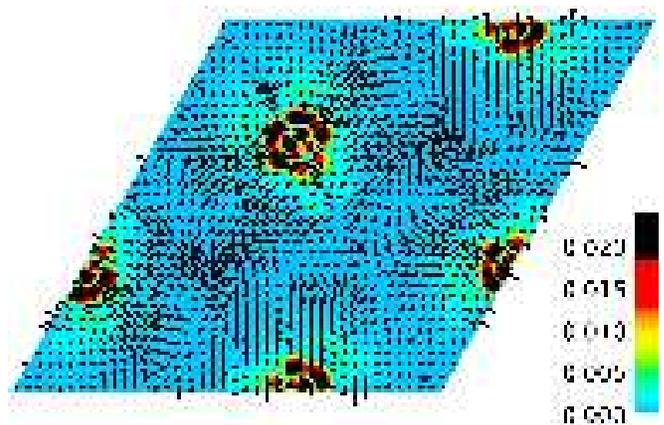}}
\caption{\label{fig:Cr2} ${\bf m}({\bf r}) \times {\bf B}_{\rm x}({\bf r})$
for an unsupported Cr-monolayer, in the same plane as Fig. 1, obtained using 
the EXX functional. Arrows indicate the direction and information about
the magnitude (in atomic units) is given in the colour bar.}
\end{figure}

Another appealing property of the EXX functional that could have consequences
in future time-dependent extensions is the non-vanishing cross product of the
magnetization density and EXX ${\bf B}_{\rm x}$ field. 
This is interesting because the equation of motion for the spin magnetization reads

\begin{align}\label{Eq:SD}
{{d{\bf m}({\bf r},t)}\over{dt}}=
\gamma {\bf m}({\bf r},t) \times 
[{\bf B}_{\rm xc}({\bf r},t)+{\bf B}_{\rm ext}({\bf r},t)]-\nabla \cdot {\bf J}_s
\end{align}
where ${\bf J}_s$ is the spin current and $\gamma$ the gyromagetic ratio.  
In the time-independent LSDA and conventional GGA, ${\bf m(r)}$ and 
${\bf B}_{\rm xc}({\bf r})$ are locally collinear, as is clear from Fig. 1, and 
therefore ${\bf m}({\bf r}) \times {\bf B}_{\rm xc}({\bf r})$ vanishes. This 
also holds true in the adiabatic approximation of time dependent SDFT which, by 
Eq.(\ref{Eq:SD}), implies that these functionals cannot properly describe the 
dynamics of the spin magnetization. In contrast, already at the static level, 
for the EXX functional
${\bf m}({\bf r}) \times {\bf B}_{\rm x}({\bf r})$ does not vanish (see Fig. 2)
In fact, in the ground state of a non-collinear ferromagnet
without external magnetic field, ${\bf m}({\bf r}) \times {\bf B}_{\rm xc}({\bf r})$
exactly cancels the divergence of the spin current, $\nabla \cdot {\bf J}_s$, i.e. these
terms are equally important, and it is essential to have a proper description of
${\bf m}({\bf r}) \times {\bf B}_{\rm xc}({\bf r})$.
These results indicate that a time-dependent generalization of our method 
could open the way to an {\it ab-initio} description of spin dynamics.
How well this functional really performs in describing the spin dynamics remains 
a question for future investigations.

We now turn to the question of the calculation of magnetic moments of collinear 
solids with the present formalism using the EXX functional.
For the collinear magnets Fe, Co, and Ni we
find moments of 2.71 $\mu_B$ (2.12 $\mu_B$), 1.77 $\mu_B$ (1.71 $\mu_B$), 
and 0.50 $\mu_B$ (0.55 $\mu_B$) respectively, where the LSDA results are indicated 
in brackets. Surprisingly, a previous OEP calculations \cite{kot97,kot98} found much 
larger moments of 3.40 $\mu_B$, 2.25 $\mu_B$ and 0.68 $\mu_B$ respectively.
This discrepancy may be attributed to the following facts: first,
the previous calculations used the atomic sphere approximation for the scalar
potential and the atomic moment approximation for the magnetization. In our work
there is no shape approximation for the scalar potential and the magnetization
is treated as an unconstrained vector field. 
Second, and more important,  in the present work a coupled set of equations is solved
to numerically invert the response function. This has the advantage of 
automatically including the response of the system to a constant magnetic field 
which is important for spin-unsaturated systems. This response needs additional 
treatment in the case where a decoupled set of equations is used and the response 
is inverted in a constant-free basis, as done in all past calculations \cite{kot97,kot98}.
We suspect that this is the major reason for the present discrepancy.

To conclude we have presented a generalization of the widely used OEP
equations for non-collinear magnetic systems. The resulting method does
not need any assumption of local collinearity for ${\bf m}({\bf r})$ 
and ${\bf B}_{\rm xc}({\bf r})$, and therefore extends {\it ab-initio} approaches
to non-collinear magnetism substantially beyond the LSDA. In particular, a 
time-dependent extension of the non-collinear OEP method naturally leads to 
a new and promising {\it ab-initio} approach to describe spin dynamics.

Finally, we note that since the formalism presented here treats KS wavefunctions 
as spinors, it can be used in conjunction with spin-orbit coupling.
In particular, in $f$-electron 
systems both spin-orbit coupling and the exchange field are of crucial importance, 
where the latter is well known to be poorly treated by LSDA/GGA. Hence, the 
present work opens new interesting routes for future extensions.

We acknowledge the Austrian Science Fund (project P16227), the EXCITING 
network funded by the EU (Contract HPRN-CT-2002-00317), NoE NANOQUANTA Network 
(Contract NMP4-CT-2004-50019), Deutsche Forschungsgemeinschaft and  Swedish 
Research Council for financial support.

\end{document}